# Phases and Time-Scales of Ignition and Burning of Live Fuels

Hamid Fazeli [a], William M. Jolly [b], David L. Blunck [a]

[a] School of Mechanical, Industrial, and Manufacturing Engineering, Oregon State University, Corvallis, OR 97331, USA.

[b] USDA Forest Service, Missoula Fire Sciences Laboratory, 5775 Highway 10 West, Missoula, MT 59808, USA.

**Abstract**

Wildland fires impact ecosystems and communities worldwide. Many wildfires burn in living or a mixture of living and senescent vegetation. Therefore, it is necessary to understand the burning behavior of living fuels, in contrast to just dead or dried fuels, to more effectively support fire management decisions. In this study, the ignition and burning behaviors of needles placed in convective heat flux were evaluated. The species included longleaf pine (*Pinus palustris*), Douglas-fir (*Pseudotsuga menziesii*), western red cedar (*Thuja plicata*), ponderosa pine (*Pinus ponderosa*), western larch (*Larix occidentalis*), pacific yew (*Taxus brevifolia*), white spruce (*Picea glauca*), and sagebrush (*Artemisia tridentate*). The ignition and burning behaviors were related to live fuel moisture content (LFMC), pilot flame temperatures, and convective heat fluxes. The different phases of ignition and burning were captured using high-speed imaging. In general, four burning phases can be observed: droplet ejection and burning, a transition phase, flaming combustion, and smoldering combustion. Ejection and subsequent burning of droplets can occur prior to sustained flaming ignition only in live fuels. For some species (e.g., longleaf pine, ponderosa pine, white spruce) droplet ejection and burning can reduce ignition times relative to dried fuel with lower LFMC. In general, the transition phase tends to take longer than the flaming and droplet phases (when these occur). During the transition phase, the fuels are heated and pyrolysis occurs. Time-scales to ignition and the different phases of ignition and burning vary more among live fuels than dead and dried fuels. This conclusion indicates that other parameters, such as chemical composition and structural morphology of the fuel, can significantly influence the burning of live fuels.

**Keywords:** *live fuel, foliage fuel, moisture content, droplet burning, high-speed imaging*

1. **Introduction**

    Better prediction and management of wildland fires are essential goals for reducing their destructive impacts. Wildland fire models can be valuable tools to predict fire spreads [1]. One explicit limitation of many fire models is that they treat live fuels (i.e., live) as dead fuels that are wet [2-7]. This assumption is problematic because the storage of moisture differs between live and dead fuels. Equally important, the chemical compositions of live, dead, and dried fuels can be different [1]. Unfortunately, limitations in the fire community's understanding of ignition and fire spread among live fuels have contributed to inaccurate models of wildfire propagation [8]. Hence,

a key to enabling more accurate modeling of the burning of live fuels is a better understanding of the physics associated with ignition and fire spread among live fuels [9, 10].

Live fuel moisture content (LFMC) plays an influential role in the ignition and burning behavior of fuels [11-23]. As a result, many studies have investigated the influence of LFMC on the flammability of fuels. As expected, increasing the moisture content and thickness of foliage increases the ignition time of live fuels [17-21]. Moisture evaporating from fuels can impact burning by diluting flammable pyrolyzates (i.e., gas around the live fuel), absorbing thermal energy, possibly altering combustion reactions, and ultimately delaying the time to ignition in dead and live fuels [11, 17-22]. In some instances, live fuels have displayed blistering on the surface of the fuel, droplet ejection, or bursting of the structure while burning [19-22]. Note, however, that changes in ignition phenomena caused by differences in species composition, moisture content, or environmental conditions have not been established.

Moisture storage differs between dead and live fuels. Dead fuels store moisture inside the fiber structure of the foliage as an inter-structure liquid and vapor. In live fuels, most of the water is stored in the xylem, and the remaining moisture is stored in cells. During the heating of live fuels, liquids, including water and extractives, begin to evaporate inside the cells. Eventually, the gases exit from the foliage and are released [11, 25]. Released gases can differ depending on the heat flux and foliage cellular structures [11]. The first form of gas release is caused by rupturing of the cell walls as the pressure increases within. The gas is consequently released through tiny pores on the foliage. The second form of gas releasing can be observed as an "eruptive jetting" or "micro-explosion," often with an audible phenomenon. In some species, eruptive jetting impacts the burning of live fuel and increases its flammability [25, 26].

Previous studies have investigated the relationship between LFMC, ignition, and burning of live fuels. However, LFMC might not be the only parameter for predicting the occurrence of some fires [18, 19, 27]. Other parameters that affect the burning of live fuel, such as foliar (leaf) chemistry [29-33] or heat flux rates [34-37], should be investigated [10, 28]. Of note to this work, two modes of heat transfer are dominant in the spread of wildfires. Radiation heat transfer serves as a pre-heating process to increase the surface temperature of live fuels, while convection further increases their temperature [33, 34]. A wide range of heat fluxes can occur in wildland fires [36]. As a result, the ignition sensitivity of dead and live fuel exposed to various heat fluxes is a key parameter that must be considered when predicting the ignition and burning of live fuels.

With this motivation and background, the overall goal of this work is to better understand the ignition process of live fuels. This goal is accomplished through five specific objectives. First, to identify and quantify the time-scales associated with the various phases leading up to and including the burning of live fuels. Second, to determine how the phases of ignition and burning change with drying of fuels for different species. Third, to better understand how moisture content impacts the burning of live and dead fuels. Fourth, to learn how the ignition sensitivities of dead and live fuels are affected by changing the heat fluxes. Fifth, to compare how different species of vegetation ignite and burn. It is expected that knowledge gained in this study can be used by the fire community to better understand the ignition and burning of live fuels and can be used to improve physics-based fire behavior models.

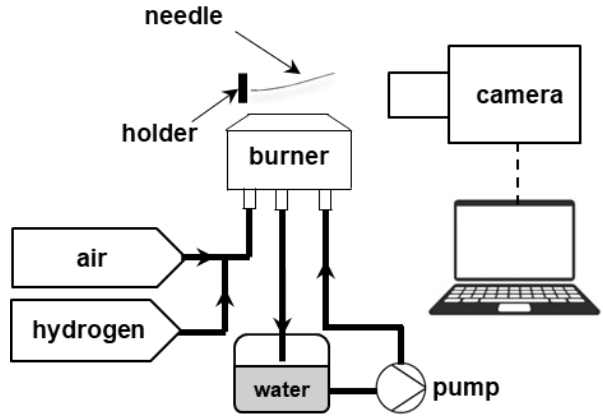

**Fig. 1** Experimental arrangement for studying ignition and burning of species for various connective fluxes.

## 2. Experimental approach
### 2.1 Experimental arrangement

Individual needles were suspended above a flat flame burner, and the resulting ignition process was recorded using a high-speed camera during the experiments. Figure 1 shows the arrangement, which consisted of flow controllers, a flat flame burner (FFB), a high-speed camera, and a cooling pump. The flat flame burner (i.e., McKenna) [38] consists of a porous plug 6 cm in diameter. The burner was operated with a mixture of hydrogen/air at an equivalence ratio (Φ) of 1.24. This Φ allowed the desired temperatures (and convective fluxes) to be obtained in a fuel-rich environment. A fuel-rich environment is expected for many conditions where flames are near vegetation. The MKS flow controller (model 247D) was used to adjust the volumetric flow rate of hydrogen and air for the burner. The MKS was calibrated using the Alicat mass flowmeter (model M250SLPM-D/5M).

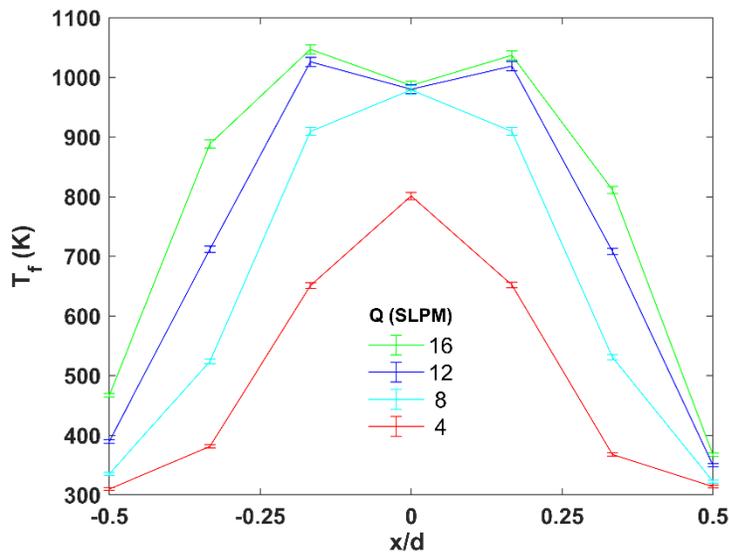

**Fig. 2** The distribution of the flame temperature ($T_f$) above the burner for different heating cases. The overall equivalence ratio (Φ) was 1.24.

The tests were conducted with the individual needle held 2 cm above the burner surface. Figure 2 shows the temperature ($T_f$) distribution across the burner at this height for different flow rates or reactants. A type-K thermocouple was used to collect the measurements. Changes in flow rates allowed the convective fluxes to be varied. The mean flame temperature, which was used to estimate convective fluxes to the foliage, was based on the average of the three middle measurement locations. The mean flame temperatures that were studied were 1025, 1010, 930, and 700 K.

A Phantom V711 high-speed camera was used to capture images of the ignition process at a frequency of 1000 frames per second (fps) and an exposure time of 400 (μs). High-speed images were analyzed to identify different processes and their time-scales leading up to ignition. The camera software was used while analyzing images.

### 2.2 Fuel collection and analysis

Ignition and burning behavior of seven species of conifer trees and one shrub were evaluated during tests, as listed in Table A.1. The specific species considered included longleaf pine (*Pinus palustris*), ponderosa pine (*Pinus ponderosa*), Douglas-fir (*Pseudotsuga menziesii*), western larch (*Larix occidentalis*), western red cedar (*Thuja plicata*), white spruce (*Picea glauca*), pacific yew (*Taxus brevifolia*), and sagebrush (*Artemisia tridentate*). These species were considered because of their significance to wildfires, their availability for harvesting, and their contrast in chemical composition. All coniferous species were harvested in the regions near Corvallis, Oregon (44.5638° N, 123.2794° W), and sagebrush was harvested near Bend, Oregon (44.0429° N, 121.3334° W) in the United States of America. Species were harvested and burned from September to February. The live fuel samples were kept in a plastic bag to prevent water and volatile loss.

Three types of samples were tested: live fuels, live fuels that were allowed to dry (referred to as dried fuel), and dead fuels. The live fuels were collected and burned within a day of being harvested, with the exception of sagebrush, which was burned within a week after harvesting. Typical moisture contents were 120-155%. Dried fuels were dried in the air for 30 days to achieve moisture contents in the range of 5-120%. Dead fuels were fallen leaves/needles collected near the selected shrub/trees and dried to reach a moisture content close to 0. It is noted that the samples were typically collected from the same trees. As such, the chemical composition of the samples for a species of shrub/tree is expected to be similar between tests other than seasonable variability.

The oven-drying method was used to determine the LFMC [41]. The samples were placed in an oven at 103°C for four days to dry. The percentage of LFMC as per the dry weight of the sample was calculated using the relationship,

$$\text{LFMC}(\%) = \frac{\text{weight of the liquid in the sample}}{\text{the dry weight of the sample}} = \frac{W_{\text{wet}} - W_{\text{dry}}}{W_{\text{dry}} - W_{\text{cont}}} \times 100\%, \quad (1)$$

where $w_{\text{wet}}$ is a wet sample, $w_{\text{dry}}$ is the weight of a dry sample, and $w_{\text{cont}}$ is the weight of the container.

### 2.3 Estimation of convective heat flux

The convective heat flux to the needle/leaf,

$$q_{\text{conv}} = h(T_m - T_s), \quad (2)$$

was estimated for the different flow conditions and needle geometries. Here *h* is the convective heat coefficient, $T_m$ is the mean flame, and $T_s$ is the initial temperature of the fuel prior to insertion (i.e., 20°C).

Nusselt number (*Nu*) correlations for a cylinder or a flat plate were used to estimate convective heat coefficient (*h*), depending on the geometry of the needle. The correlation for a cylinder in a cross-flow [39],

$$Nu_D = \frac{hD}{k} = 0.3 + \frac{0.62 Re_D^{1/2} Pr^{1/3}}{[1+(0.4/Pr)^{2/3}]^{1/4}} \left[1 + \left(\frac{Re_D}{282000}\right)^{5/8}\right]^{4/5} \qquad PrRe_D > 0.2, \qquad (3)$$

was used to characterize convective heat transfer for the needle-type species, including longleaf pine, Douglas-fir, ponderosa pine, western red cedar, white spruce, western larch. *k* is the thermal conductivity of air, *D* is the characteristic diameter of the needle, *Pr* is the Prandtl number of the air, and $Re_D$ is the Reynolds number of gas flow around the needle. The width of a needle was assumed as *D*. The shapes of the needles are not truly cylindrical, but this geometry seemed most representative for heat transfer correlations.

The correlation for a flat plate was determined by Hilpert and Reiher [40]:

$$Nu_L = \frac{hL}{k} = 0.231 Re_L^{0.731} \qquad (4)$$

was used to estimate convective fluxes through leaves of pacific yew and sagebrush. Here *L* is the projected width of the flat plate (i.e., leaf) perpendicular to the freestream. The *Re* of the flow was less than that evaluated for the correlation ($6.3 \times 10^3$ to $2.36 \times 10^4$). However, no other suitable correlation valid for the *Re* of this study (i.e., between 0.3 to 18.8) was found. Thus, the relationship (4) was used to approximate the Nusselt number for flow around the aforementioned species. The ranges of convective heat fluxes evaluated were between 5 to 95 (kW/m$^2$), as reported in Table A.2. In order to validate the estimated heat fluxes, the range of heat flux compared to other works [34, 36] are within the acceptable range between 5-100 kW/m$^2$. Four different heat fluxes are estimated for the four different gas temperatures above the burner, as listed in Table A.2 for each species.

## 1. RESULTS AND DISCUSSIONS
### 1.1 Multi-phase of burning in live fuels

Ignition and burning of the samples can occur across several phases, as identified from the high-speed images. The occurrence and duration of these phases vary depending on the species, heat flux, and moisture content. The phases can include droplet ejection/burning, transition, flaming, and smoldering combustion. The characteristics of each phase are now briefly described and discussed.

#### 1.1.1 First phase: Droplet Burning

Droplets form on the surface of the foliage as samples are inserted into the flame, and, in some instances, the droplets are ejected from the surface. As will be discussed shortly, the evidence shows that ejected droplets burn, particularly after leaving the surface. Fig. 3a shows an example of droplets on the needle surface and all three panels of Fig. 3 show droplets that have ejected and are burning. The size of the droplets in the air in Fig. 3.b is estimated to be <<< 0.5 mm.

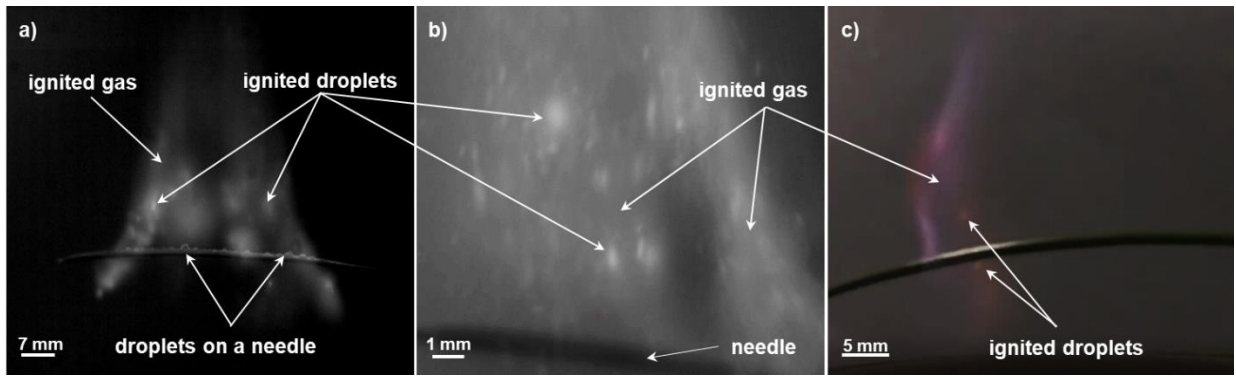

**Fig. 3** High-speed and color images of droplet burning in (a) longleaf pine (b) Douglas-fir (c) ponderosa pine.

Flames can be observed during the droplet burning phase, as shown in the color image of Fig. 3.c. It is presumed that the surface droplets burn, but this is difficult to verify. It is noted that these droplets are not always observed, depending on the heat flux, species composition, and moisture content of the fuel.

Physically, droplets leave the foliage as the heating causes the intercellular liquid to heat, partially vaporize, and leave the plant. The liquid consists of a mixture of water and extractives [33]. Openings in the epidermis of the needles (via stomata) allow droplets and vapor release [11], although presumably, the epidermis may rupture in some conditions and provide a second avenue for moisture release.

A mist of saturated water-sugar solution and water was sprayed above the burner to evaluate whether the droplets observed were actually burning or simply reflecting light. The water-sugar solution was a simple surrogate for fluids ejected from foliage. A pressure atomizer was used to inject spray into the flame. Representative images from the experiments with a saturated water-sugar solution and pure water are shown in Fig. 4. The droplets (< 1 mm in diameter) with saturated water-sugar solution are visible (see Fig. 4a), similar to the droplets observed in Fig. 3. In contrast, no droplets were visible with just water (see Fig. 4b). The results from this experiment are evidence that the droplets observed above the needles are burning and not just visible because of reflected light.

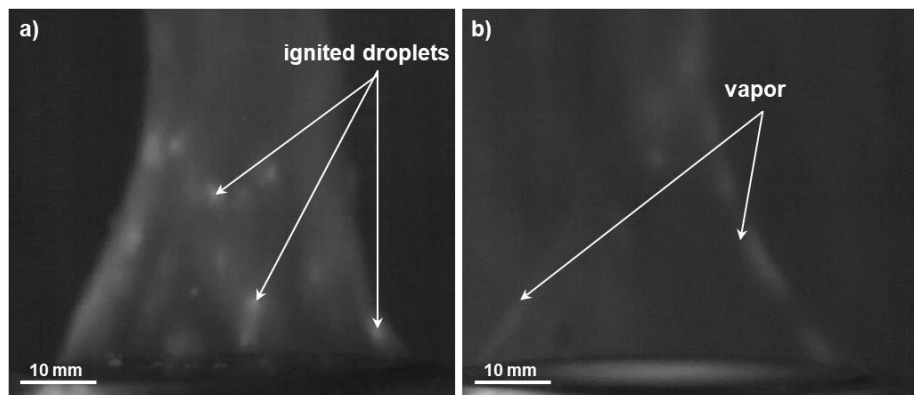

**Fig. 4** High-speed images of droplet burning of (a) saturated water-sugar solution (b) vapor of pure water.

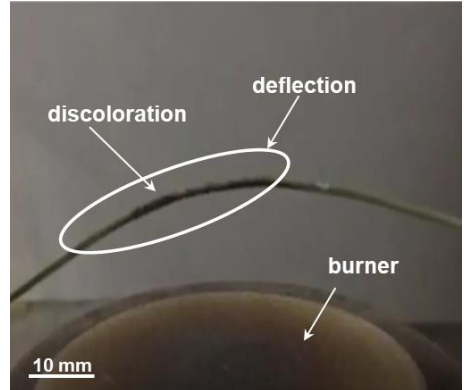
**Fig. 5** The transition phase with discoloration and deflection in the ponderosa pine needle.

The significance of droplet ejection is that this process distributes both fuel and moisture away from the foliage. As a result, the overall burning time can be decreased under some conditions, as discussed in Section 3.3.

### 3.1.2 Second phase: Transition

The visible discoloration shown in Fig. 5 is evidence of pyrolysis during the transition phase. Long-chain carbon molecules such as lignin, cellulose, and hemicellulose are converted into pyrolyzates [42]. Ignition indicates the end of the transition phase. In some experiments, eruptive jetting occurred during transition and/or droplet burning phases. Eruptive jetting refers to the rapid ejection/burning of the gases from a needle/leaf to the surroundings [35, 36]. An example of this process is shown in Fig. 6. Eruptive jetting is not considered a separate phase in this work as it may overlap the transition and/or droplet phases. The jets produced by this process lack discrete droplets and tend to be larger than droplet ejections, allowing the two to be differentiated. Figs. 6a and 6b provide examples of gases ejected from the needles. Eruptive jetting can occur in single or multiple directions. Burning of the ejected gases can occur, but the lower light intensity shown in Fig. 6c indicates that burning does not always happen. Physically, eruptive jetting is expected to result from rupture of the cell walls due to gas pressure accumulating in the cells as liquids vaporize.

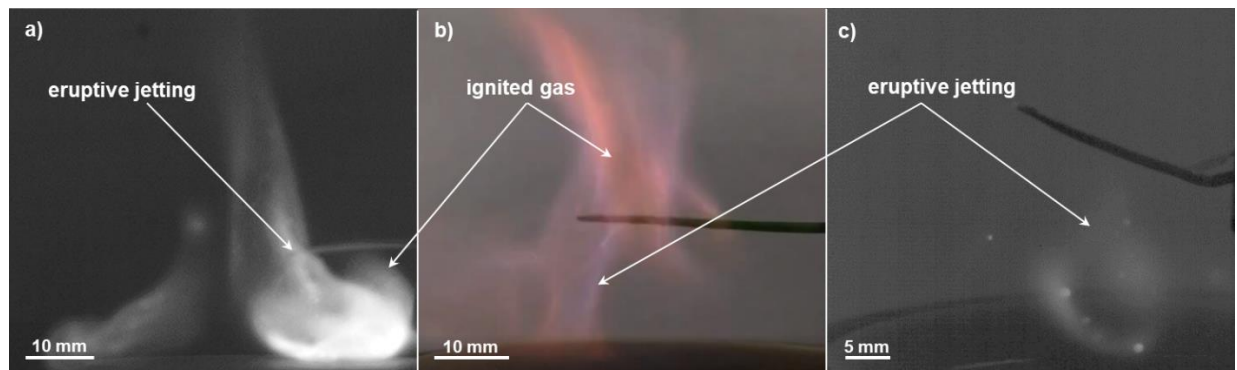
**Fig. 6** High-speed and color images of micro-explosion (eruptive jetting) in (a) Douglas-fir (b) western larch (c) pacific yew.

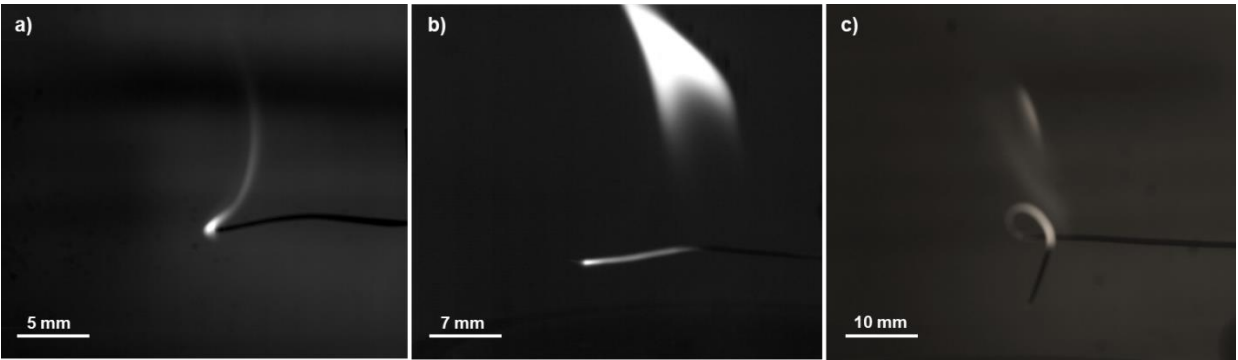

**Fig. 7** The flaming combustion process in different coniferous species; (a) ignition in Douglas-fir, (b) flaming in western larch, (c) burnout in longleaf pine.

### 3.1.3 Third phase: Flaming Combustion

The start of the flaming combustion phase is marked by a sustained flame on the needle/leaf. Examples of flaming combustion are shown in Fig. 7. The phase starts with ignition (panel a), continues into flaming (panel b), and ends with burnout (panel c). Ignition typically begins near the tip of the sample and spreads from there. The apparent area of the flames tends to increase (see Fig. 7b) as heat from the flame increases the rate at which pyrolyzates are released. The flames continue until the volatile gas concentration reaches the flammability limit and the flame disappears (i.e., burnout), as shown in Fig. 7c. A difference was observed between live and dead fuels in this phase – dead fuels display greater maximum flame height and visible flame intensity. Presumably, the water vapor released during the burning of live fuels dilutes the concentration of pyrolyzates, leading to lower temperatures.

### 3.1.4 Fourth phase: Smoldering Combustion

Reactions often continued as smoldering combustion after the flaming phase. As shown in Fig. 8, there was no flame, but burning continued, as evidenced by continued visible light emissions from the surface of the needle or smoke. The end of smoldering combustion is the "extinction" point when the burning stops. Ignition occurred without flames in some species (e.g., white spruce, pacific yew, and sagebrush). Ignition was identified by smoke or light emitted by the foliage.

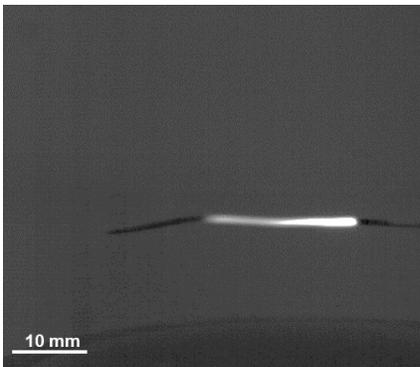

**Fig. 8** An example of smoldering combustion in the western larch needle.

## 3.2 Surface temperature for different phases of burning

Temperature measurements were collected from heated longleaf pine needles to better understand the influence of temperature on the various phases leading up to and following ignition. These measurements were matched with a visual assessment of the ignition phases in order to correlate the two. Longleaf pine needles were selected because their relatively thick structure allowed the thermocouples (type-K) to be bound to them so that the tip was generally in contact with the surface. With this approach, the temperatures are considered qualitative estimates of the surface temperatures. Nonetheless, a correlation between the different phases and the temperature of the needles is evident.

Figure 9 shows the temperatures of the needles as they were heated and transitioned through the four ignition and burning phases. The first phase (droplet burning) occurred at a temperature of roughly 80°C to 100°C. This temperature range shows that droplet burning and water vaporization happened simultaneously. After the droplet burning phase terminated, the temperature increased during the transition phase. Subsequently, flaming combustion of the samples was observed near temperatures of 310 °C to 350 °C.

The estimated temperature of the needles continued to increase until near 600 °C. The smoldering phase then occurred with relatively few changes in the estimated surface temperature. The surface temperatures of longleaf pine were compared to the results of Dietenberger et al. [42], who reported surface temperatures of the same fuel as it was heated using cone calorimetry. The variation of surface temperatures in Fig. 9 follows the surface temperatures in the results in [42]. Surface temperatures in the current and former work increase after ignition to reach 600 °C. Finally, the variation of surface temperatures shows an approximately constant surface temperature of about 700 °C in Fig. 9 in the smoldering phase, similar to the results in [42].

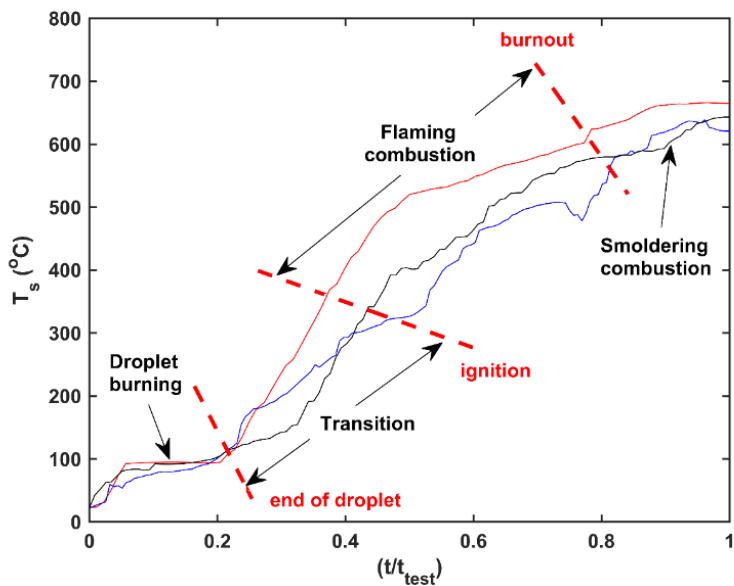

**Fig. 9** The variation of surface temperature ($T_s$) in three samples of longleaf pine with 125% moisture content.

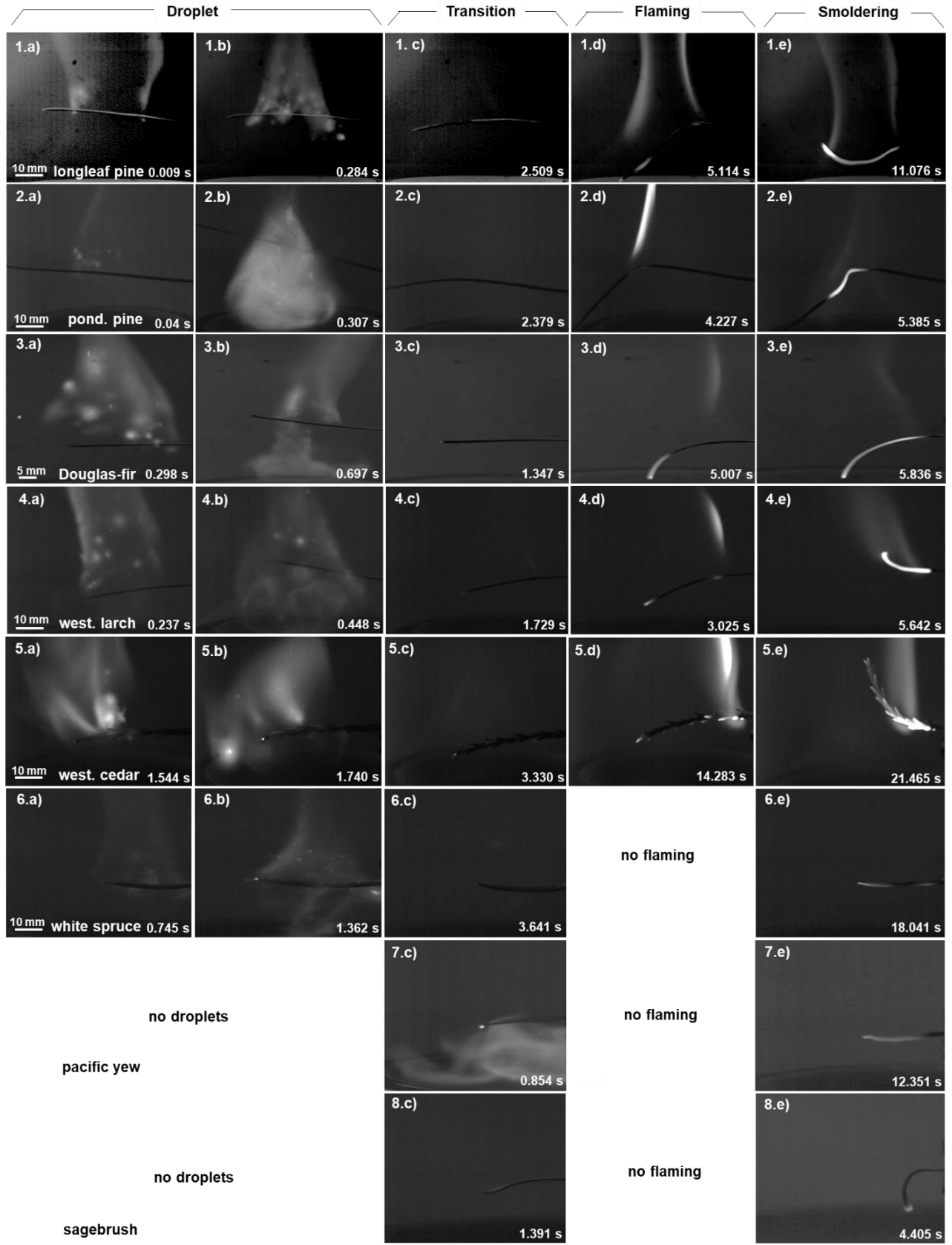

**Fig. 10** Multi-phase of ignition and burning in live fuels. Time 0 (s) refers to when the needle was placed above the burner.

### 3.3 The contribution of multi-phases of ignition and burning

Representative high-speed images of the ignition and burning processes of the eight species studied in this work are shown in Fig. 10. Each row of Fig. 10 includes images illustrating the different ignition and burning phases. More broadly, the four burning phases discussed earlier were observed for longleaf pine, ponderosa pine, Douglas-fir, western larch, and western red cedar needles. For white spruce, all of the stages were observed except for flaming combustion. Transition and smoldering combustion occurred for pacific yew and sagebrush, but the other phases did not. Specific aspects of the ignition process are now highlighted for the various species.

In longleaf pine, ponderosa pine, Douglas fir, western larch, western red cedar, and white spruce, ejection and burning of droplets started almost immediately (within 10 ms) upon the sample being exposed the flame. Figs. 10.1.a through 10.6.a, show the formation of burning droplets on the needle. The droplets formed on the needle's surface then spread along its length. In some instances, eruptive jetting occurred during the droplet burning phase, as seen in ponderosa pine (10.2.b), Douglas-fir (10.3.b), western larch (10.4.b), and white spruce (10.6.b).
The transition phase was considered to begin when there was no indication of droplet burning, as shown in Figs. 10.1.c through 10.8.c. Note that pacific yew still exhibited eruptive jetting in this phase. After starting the transition phase, needle discoloration indicated pyrolysis had begun. The end of this transition phase was identified by flaming ignition as shown in Fig. 10.1.d to 10.5.d or smoldering ignition (visible as smoke) as shown in Fig. 10.6.d to Fig. 10.8.d. Five species showed flaming combustion, while white spruce, pacific yew, and sagebrush did not. This observation is notable because larger masses of these species display flaming combustion. It seems probable that the concentration of pyrolyzates near the samples was insufficient for flaming ignition. More broadly, the variable reactions of the different species illustrate that both the mass and composition of a species can be essential factors contributing to flaming combustion.

Smoldering was identified in all species, as shown in Fig. 10.1.e through Fig. 10.8.e. In white spruce, pacific yew, and sagebrush, the smoldering happened directly after the transition phase without any intervening flame. It is assumed that the release rate of pyrolyzates is insufficient for these species to cause a flammable mixture to form.

A brief discussion is provided to summarize what happens at needles/leaf scales during the heating process as fuels are heated, intercellular liquids are heated, and vaporized. Live fuels generally have a moisture content per unit dry weight of ~100%, which means half of the fresh foliage mass is water. The water inside the foliage is in solution with other compounds. Typically, gas exchange occurs through tiny openings in the foliage called stomata. The temperature of the foliage remains relatively low (e.g., 100°C) during heating because of the relatively large amount of energy required to raise the water temperature and subsequently vaporize it. For thermally thin fuels, the temperature of the foliage increases after the moisture leaves the foliage. Pyrolysis and subsequent ignition ensue after the temperature of the foliage is sufficient. Further, the external waxy covering of the foliage also evaporates, creating more pathways for vapor to escape. Droplet or eruptive jetting occurs when the mass flux of vapor from the foliage exceeds what can transport through the stomata or the cuticles. The pressure within the foliage increases until the cellular structures rupture, and the vapor causes residual liquids to be rapidly expelled and burn.

### 3.4 Influence of LFMC and heat flux on ignition time and behavior

Figures 11 and 12 show the variation in ignition time relative to heat fluxes and LFMC for the various species. These results are provided to better understand the influence of heat transfer and fuel moisture content on the time to ignition and the different ignition phases. The LFMC varied among freshly cut, partially dried, and dead samples. For reference, the range of LFMC in live and dried fuels was between 120-155% and 10-116%, respectively. The moisture content for dead fuel was lower than 5%. The estimated convective heat flux range was within 10-95 (kW/m$^2$) for longleaf pine, ponderosa pine, Douglas-fir, and western larch, as shown in Fig. 11.d. The heat flux range was within 5-55 (kW/m$^2$) for western red cedar, white spruce, pacific yew, and sagebrush, as shown in Fig. 12. Note that the same temperatures and exhaust velocities above the burner were used for the experiments; the heat flux varied because of differences in the geometries of the needles. The results are the average ignition times of four tests for each species at a particular moisture content.

The filled symbols '●' indicate conditions whose ignition process contains droplets burning, while non-filled symbols '○' are conditions where no droplets were observed. Representative precision uncertainty bars (95% confidence) are included.

Ignition times decrease with decreasing LFMC for all species, as shown in Figs. 11.e through 11.h and Figs. 12.e through 12. H. Decreasing the LFMC reduces the energy required to evaporate the moisture and increase foliage temperature to a level sufficient for pyrolysis to occur. Douglas-fir and western larch usually had shorter ignition times than longleaf pine and ponderosa pine, even at similar (i.e., within 5%) moisture contents. This sensitivity indicates that discrepancies in ignition times can occur because of differences in foliar chemistry.

Droplet burning (represented by filled symbols in Figs. 11 and 12) appeared for the highest heat flux conditions [i.e., $T_g$=1025 K and $T_g$=1010 K heating cases] and for the highest moisture contents. This observation shows that the moisture within the foliage and the heating rate must be sufficient to cause droplets to be released. If the moisture content or heating rate is insufficient, moisture is released more gradually from the foliage. The only dried fuel that showed droplet burning in its samples was ponderosa pine with LFMC of 116%.

For some species, droplet ejection and burning tend to reduce the ignition times. This reduction in ignition times is evidenced by the inflection in the results (see Figure 11 panels e and f; Figure 12 p. f). Anecdotally, a reduction in ignition time occurs for species with the greatest ignition sensitivity to the LFMC (e.g., ponderosa pine and longleaf pine). Droplet burning, however, does not tend to reduce ignition times for species whose ignition times are less sensitive to moisture content (e.g., Douglas-fir and western larch needles).

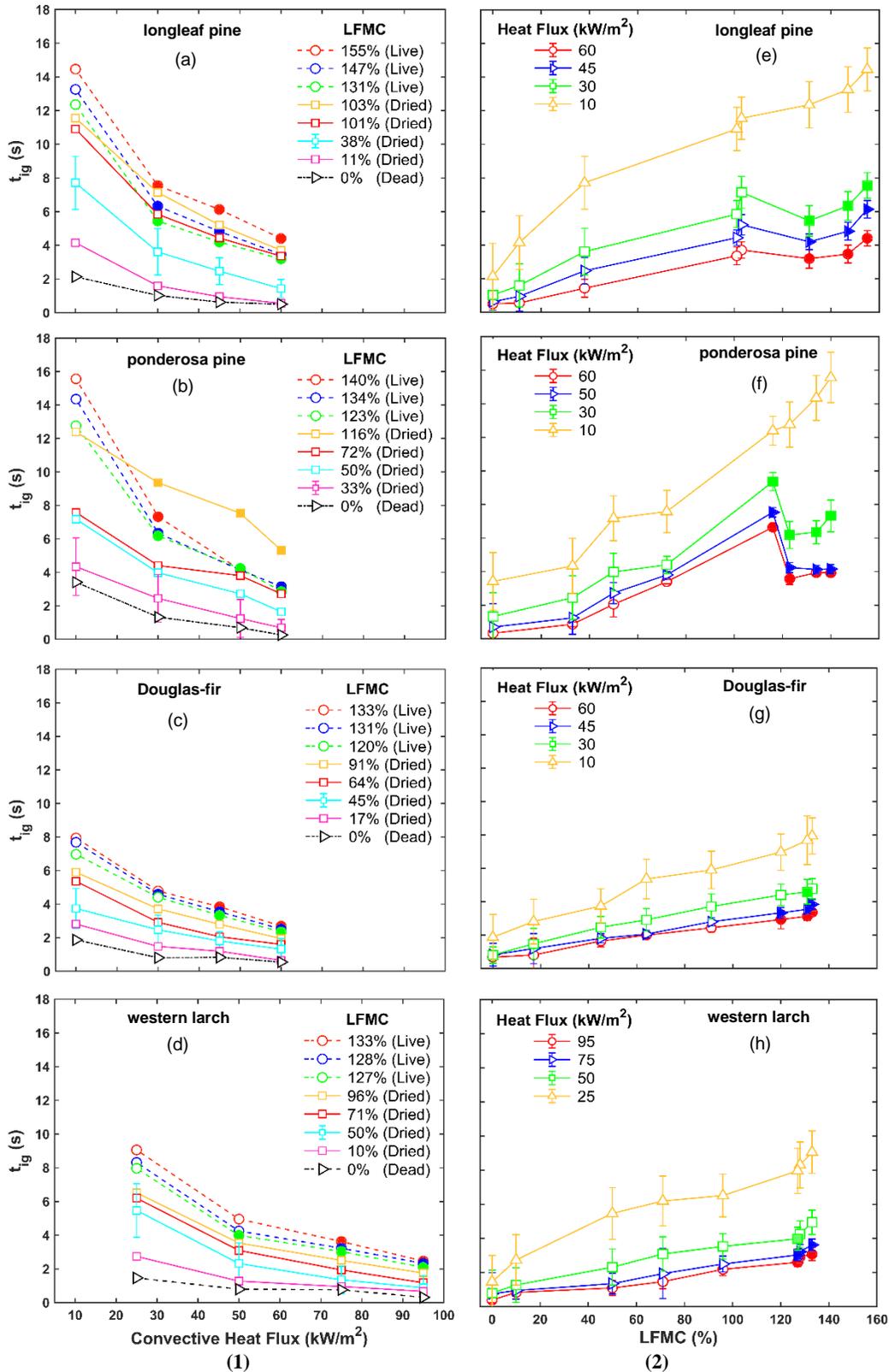

**Fig. 11** The variation of ignition time versus (1) heat fluxes and (2) live fuel moisture content: with (●) and without (○) droplet burning. Same convective conditions (i.e., velocity and temperature) were used for all tests.

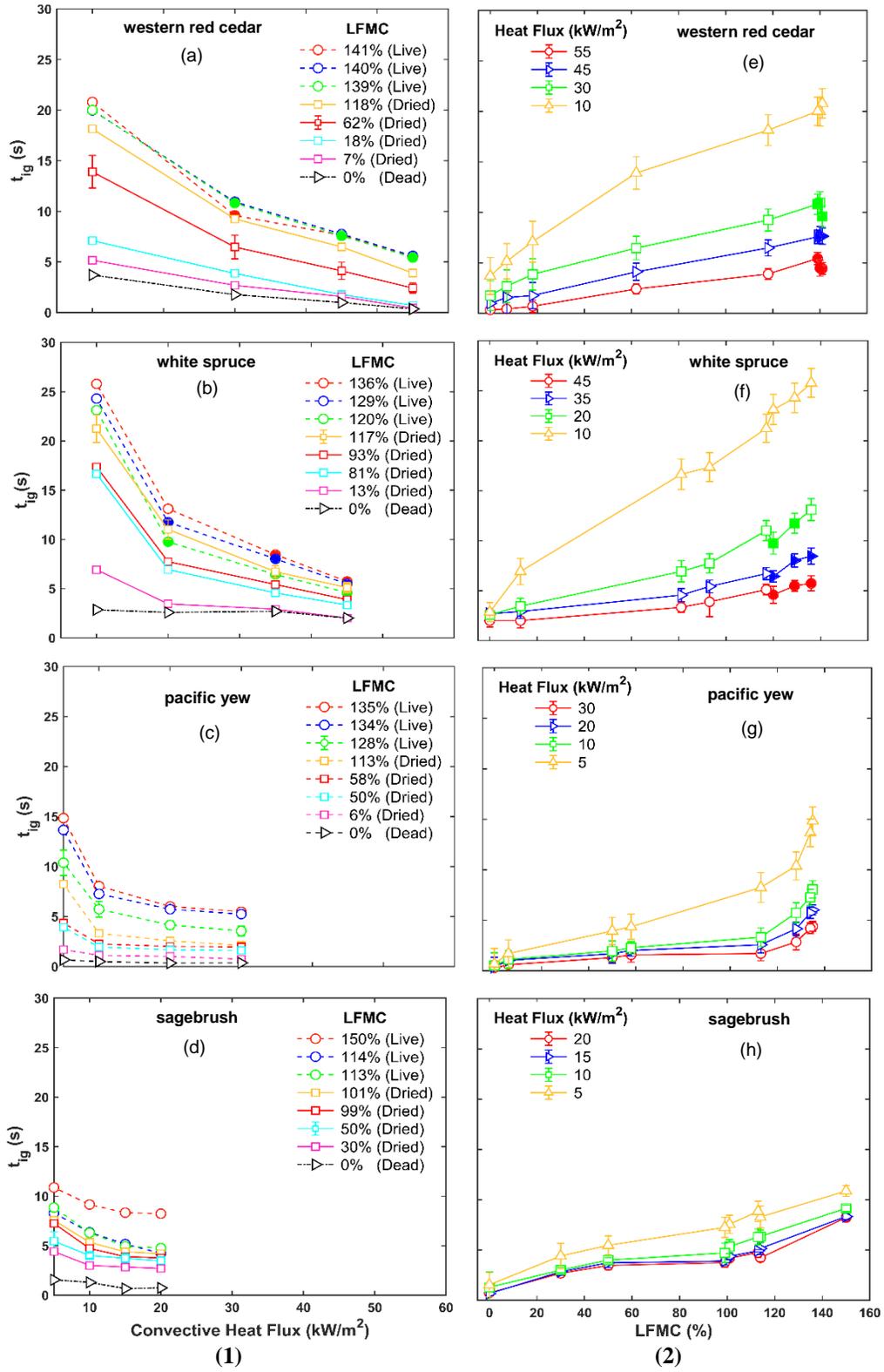

**Fig. 12** The variation of ignition time versus (1) heat fluxes and (2) live fuel moisture content: with (●) and without (○) droplet burning. Same convective conditions (i.e., velocity and temperature) were used for all tests.

### 3.4.1 Ignition sensitivity of live and dead fuels

An ignition sensitivity index (ISI) was defined to quantify the sensitivity of a given species' time to ignition between the highest and the lowest convective heat fluxes.

For needle-type:
$$\mathbf{ISI} = \frac{4 \cdot \Delta t_{ig} \Delta q''}{\rho \overline{c_P} D \Delta T_g} \tag{5}$$

while for leaf-type,
$$\mathbf{ISI} = \frac{2 \cdot \Delta t_{ig} \Delta q''}{\rho \overline{c_P} W \Delta T_g} \ . \tag{6}$$

Here, $\rho$ is the density of live fuels (360-0.540 kg/m³) and dead fuels (380-650 kg/m³) [43], $\overline{c_P}$ is the average specific heat capacity of fuels between the lowest and highest heat fluxes (i.e., the specific heat capacity of fuels depends on changing temperature) [43, 44], $\Delta t_{ig}$ is the difference between the highest and the lowest ignition times. $\Delta T_g$ corresponds to the gas temperature difference between the highest and the lowest heating cases. Ideally, surface temperatures would be used in Eqs. 5 and 6., however, $T_g$ is sufficient considering that it is known and influences surface values. Here $D$ and $W$ are either the characteristic needle diameter (Eq. 5) or the leaf thickness (Eq. 6). The factors of 4 and 2 result from simplifying the volume and surface area of the samples in the relationships. Physically the ISI represents the ratio of the heat transfer to the needle (via convection) relative to the rate that the sensible energy of a sample changes. Hence samples that take longer to heat, either through endothermic reactions or vaporization, would have a larger ISI and should take longer to ignite.

Table 1 shows the ISI for the various fuels, live and dead. The ISI values for dead fuels lie between 0.02 and 0.53, while the ISI values for live fuels change in the range of 0.08 to 2.65. The exact values for ISI are not as important as the relative values and trends. Physically, the lower ISI values show that dead fuels have lower sensitivity of ignition to the heat fluxes than live fuels. Energy absorbed by dead fuels more readily changes the sensible energy of the needle. In general, the relatively ranking of ISI values tends to be consistent for a particular species (e.g., highest ISI values for both live and dead fuels, etc.). It is plausible that some of the same consistent remain within the live and dead fuels for a particular species, thus influencing the relative ISI compared to other species. Finally, it is noted that the two live species that did not display droplet ejection (i.e., pacific yew and sagebrush) have the lowest ISI among live fuels. This sensitivity is attributed to heat from convection more readily changing the sensible energy of the needle instead of causing droplet ejection.

Table 1 The ignition sensitivity of live and dead fuels relative to the highest and lowest heat fluxes.

| Label | Species | live fuel | | | | dead fuel | | | |
|---|---|---|---|---|---|---|---|---|---|
| | | $t_{ig,H}$ (s) | $t_{ig,L}$ (s) | $\Delta t_{ig}$ (s) | ISI | $t_{ig,H}$ (s) | $t_{ig,L}$ (s) | $\Delta t_{ig}$ (s) | ISI |
| LP | longleaf pine | 4.4 | 14.5 | 10.1 | 1.13 | 0.5 | 2.1 | 1.6 | 0.17 |
| PP | ponderosa pine | 3.2 | 15.6 | 12.4 | 1.61 | 0.3 | 3.4 | 3.1 | 0.38 |
| DF | Douglas-fir | 2.7 | 7.9 | 5.2 | 0.73 | 0.5 | 1.9 | 1.4 | 0.19 |
| WL | western larch | 2.5 | 9.1 | 6.6 | 2.65 | 0.3 | 1.5 | 1.2 | 0.41 |
| WC | western red cedar | 5.4 | 20.8 | 15.4 | 2.29 | 0.4 | 3.7 | 3.3 | 0.53 |
| WS | white spruce | 5.5 | 25.8 | 20.3 | 1.04 | 2.0 | 2.9 | 0.9 | 0.07 |
| PY | pacific yew | 4.2 | 14.8 | 10.6 | 0.68 | 0.3 | 0.7 | 0.4 | 0.02 |
| SB | sagebrush | 8.2 | 10.9 | 2.7 | 0.08 | 0.7 | 1.6 | 0.9 | 0.03 |

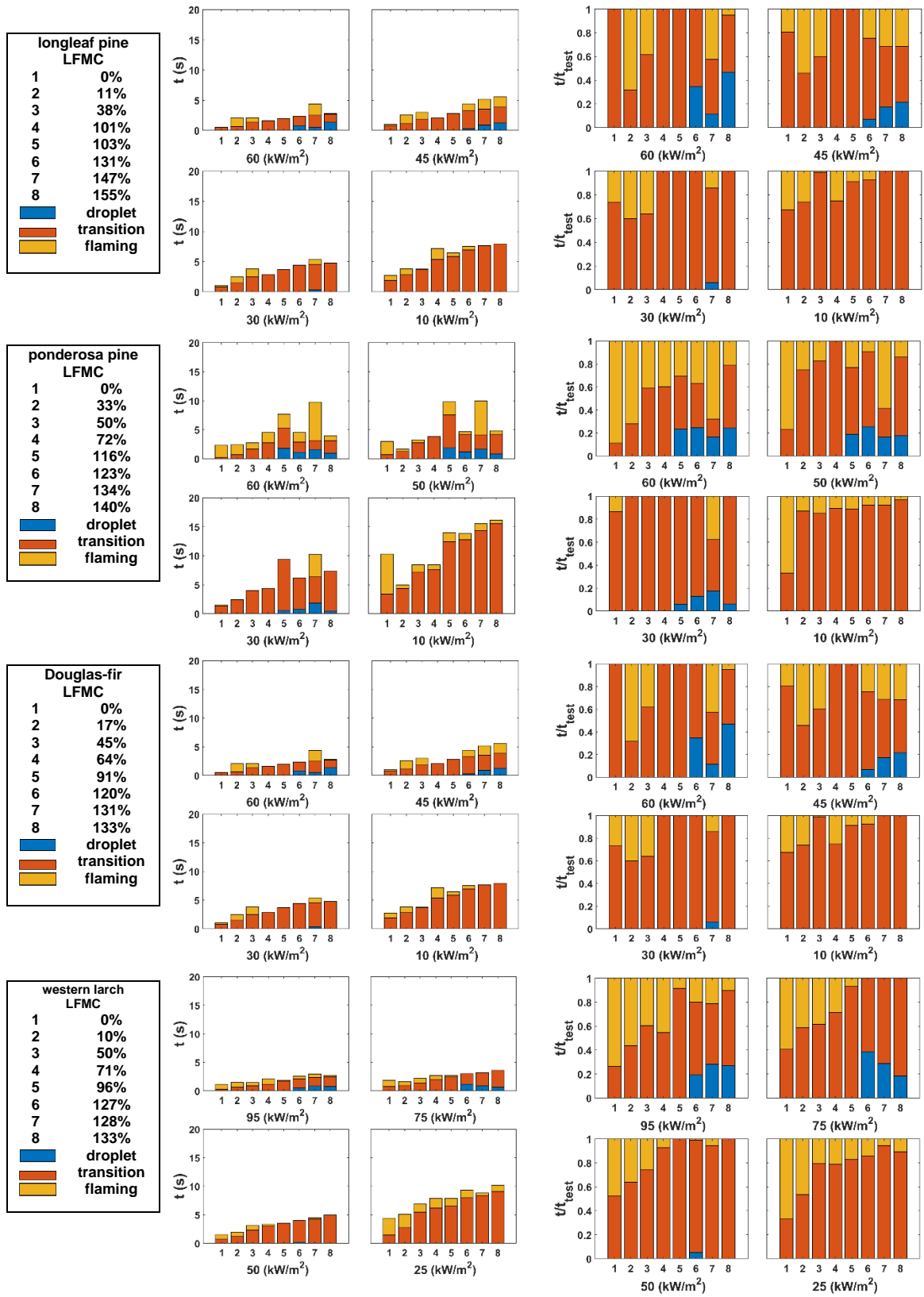

**Fig. 13** The times (column a) and normalized times (column b) associated with the different phases of ignition and burning. The results are reported for different heat fluxes.

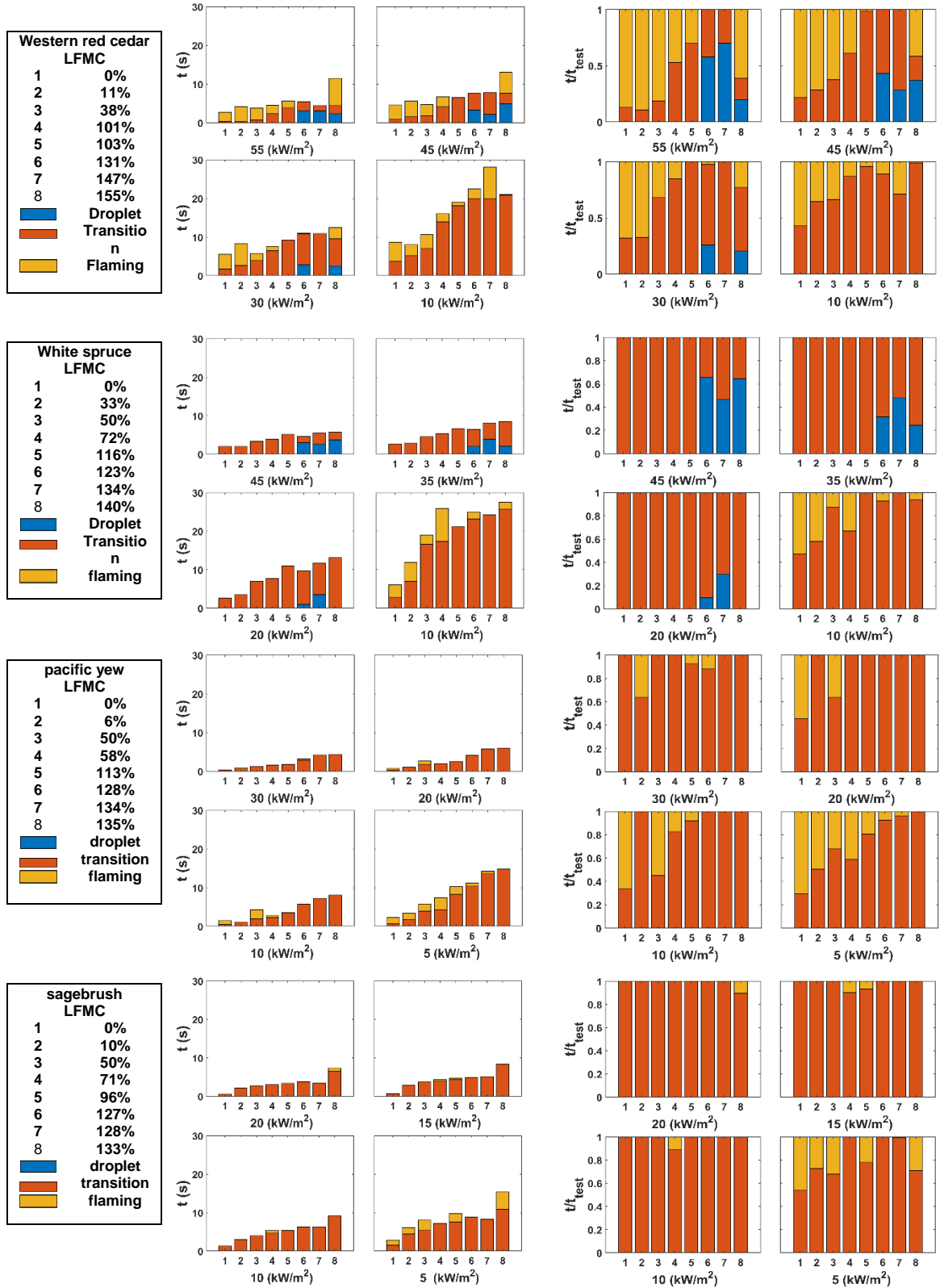

**Fig. 14** The times (column a) and normalized times (column b) associated with the different phases of ignition and burning. The results are reported for different heat fluxes.

### 3.5 The influence of multi-phases on burning

The average times associated with the different phases of ignition and burning were analyzed and are reported in Figs. 13 and 14. The left and right columns of the figures show the average specific and normalized times, respectively. Note that smoldering burning is not included in the time-scales because of challenges in identifying when smoldering concludes.

The droplet burning phase typically lasted less than 1.0 second when it occurred. The times associated with the droplet phase typically fell between 5% and 45% of the test time (i.e., inset of the needle until flaming concluded or smoldering started). In a few instances, the droplet phase took up 60% of the time. The results show that sufficient moisture needs to be present for droplet burning or ejection to occur; presumably, moisture is needed to cause cellular ruptures within the foliage. For the third heating case (corresponding to free stream $T_g = 930$ K), the occurrence of droplet burning varied for each species and was less consistent.

The transition phase occurred for all species. Pyrolyzate concentration and temperature increase until they reach ignitable conditions. The transition phase took up from 30% to 80% of the overall time, but typically more than 50% of the time from insertion to completion of flaming, as shown in Figs. 13 and 14. In a few instances, the transition phase took up 100% of the time because the samples smoldered, and the duration of the smolder was not reported. In general, the transition phase time declines with decreasing species moisture content in time and normalized time, as shown in Figs. 13 and 14. A decrease in heat flux leads to an increase in the transition phase's duration, as the greater time required to achieve an ignitable temperature would predict. In the all species time-based bar chart shown in Fig. 13 and 14, with decreasing convective heat transfer, the duration of the transition time for heat fluxes covered a significant portion of the test time in the fourth heating case (i.e., $T_g=700$ K). The results for all species show that when LFMC was larger than 30% the transition phase took a significant portion (>50%) of the test time, while in the third and fourth heating cases, the transition phase took a similar amount of time at all levels of LFMC, even in the dead fuel samples.

The flaming combustion phase did not happen in all test results shown in Figs. 13 and 14. The occurrence and duration of the flaming phase varied notably depending on the species, LFMC, and heat flux. In lower LFMC samples, the flaming phase took longer than the transition phase. For example, in the conditions with the highest convective heat flux, longleaf pine, ponderosa pine, and western larch experienced flaming. Under the same freestream conditions, flaming combustion was incomplete in other species, which shows that the possibility of flaming combustion depends on heat flux and LFMC, except in the case of white spruce, in which flaming occurred in the lowest heating cases. The duration of flaming combustion decreases from the first heating (i.e., $T_g=1025$ K) to the third heating cases (i.e., $T_g = 930$ K) in all species except pacific yew. In contrast, the fourth heating case (i.e., $T_g=700$ K) in all species shows an increase in the possibility of flaming combustion. Note that the fourth heating case (i.e., $T_g=700$ K) did not show droplet burning in all species. There may be an inverse relationship between droplet burning and increased flaming combustion in the fourth heating case.

# 4   CONCLUSIONS

This study sought to systematically identify the various phases of ignition and burning in live, dead, and dried fuels relevant to wildfires. The influence of LFMC and convective heat flux on ignition and burning behavior was examined. Four phases leading up to and including burning were observed: droplet burning, transition, flaming combustion, and smoldering combustion. The time-scale results show that the LFMC and heat flux have a critical impact on the duration of the droplet, transition, and flaming phases.

The specific conclusions of this study are as follows:

1. Ejection and subsequent burning of droplets can occur prior to sustained flaming ignition only in live fuels. The LFMC and heating flux must be sufficient to induce droplet ejection. Presumably, this is due to differences in the cellular structure and/or plant morphology of live, dried, and dead fuels.
2. Droplet ejection and burning can lead to a reduced time to ignition for live fuels relative to dried fuels with lower LFMC in some species (longleaf pine needle, ponderosa pine, and white spruce). This decreased time to ignition may be attributed to the flammable materials being more broadly distributed (as droplets surrounding the needles). However, in other species, either no droplet burning is observed (e.g., sagebrush, pacific yaw), or the time to ignition is not reduced even when droplet ejection occurs (e.g., white spruce and Douglas-fir).
3. In general, the transition phase tends to take longer than the flaming and droplet phases (when these occur). During the transition phase, the fuels are heated, and pyrolysis occurs. An exception is the conditions with higher heat fluxes (e.g., 60 kW/m$^2$) and lower moisture contents (e.g., < 30%); in these situations, flaming combustion can take longer than the transition time.
4. Both the time-scales to ignition and the different phases of ignition and burning vary more among live fuels than dead and dried fuels. This conclusion indicates that other parameters, such as chemical composition and structural morphology of the fuel, can significantly influence the burning of live fuel.
5. Ignition time is much more sensitive to the heat flux in live fuels than dead ones. This was evident from the ignition sensitivity index. The ignition sensitivity index shows that dead fuels demonstrate lower sensitivities to changes in heat fluxes.

# APPENDIX-A.1 List of plant species evaluated in this study

As depicted in Table A.1, seven species of conifer tree and one shrub were evaluated during ignition tests. The specific species considered included longleaf pine (*Pinus palustris*), Douglas-fir (*Pseudotsuga menziesii*), western red cedar (*Thuja plicata*), Ponderosa pine (*Pinus ponderosa*), western larch (*Larix occidentalis*), pacific yew (*Taxus brevifolia*), white spruce (*Picea glauca*), and sagebrush (*Artemisia tridentate*).

Table A.1 List of plant species in the experiments. Some images were obtained from reference [45].

| Scientific Name | Tree | Live Fuel | Dead Fuel | Scientific Name | Tree | Live Fuel | Dead Fuel |
|---|---|---|---|---|---|---|---|
| longleaf pine (*Pinus palustris*) | 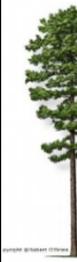 | 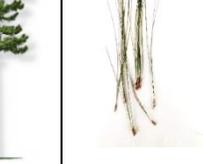 | 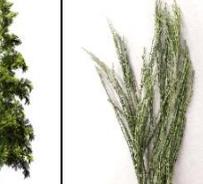 | Douglas-fir (*Pseudotsuga menziesii*) | 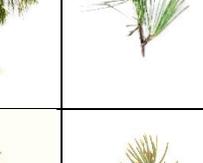 | 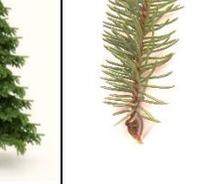 | 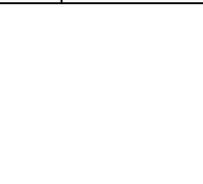 |
| western red cedar (*Thuja plicata*) | 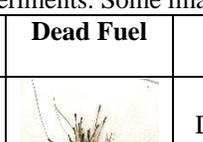 | 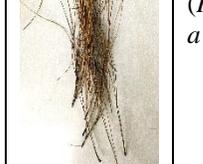 | 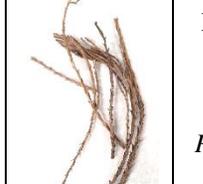 | Ponderosa pine (*Pinus Ponderosa*) | 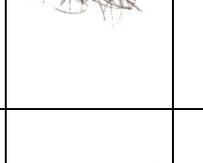 | 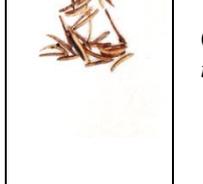 | 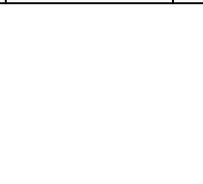 |
| western larch (*Larix occidentalis*) | 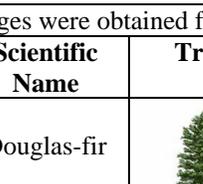 | 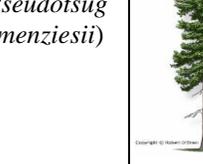 | 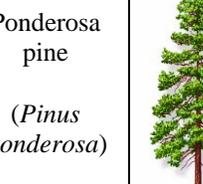 | Pacific yew (*Taxus brevifolia*) | 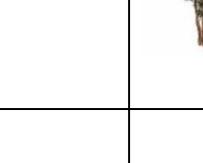 | 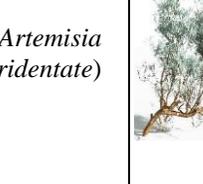 | 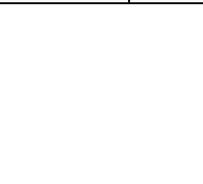 |
| white spruce (*Picea glauca*) | 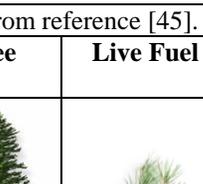 | 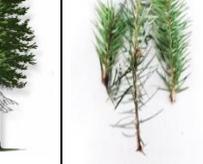 | 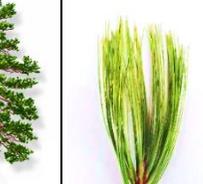 | sagebrush (*Artemisia tridentate*) | 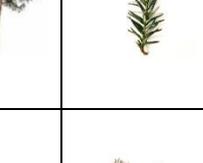 | 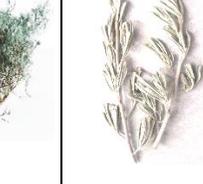 | 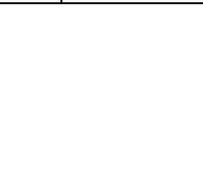 |

# APPENDIX-A.2 Different convective heat flux

The different heating conditions and the corresponding convective heat fluxes.

Table A.2 Convection heat flux approximation in heating cases for different species

| Specie | Nu correlation used | Diameter/ Width (mm) | $T_m$ (K) | $Re$ | $Nu$ | $h$ (W/m²K) | $T_m - T_s$ (K) | $q''_{conv}$ (kW/m²) |
|---|---|---|---|---|---|---|---|---|
| longleaf pine (LP) | Cylindrical (Eq. 3) | 1.4 | 1025 | 6.7 | 1.6 | 78.6 | 727 | 60 |
| | | | 1010 | 3.9 | 1.3 | 62.6 | 712 | 45 |
| | | | 930 | 2.0 | 1.0 | 45.9 | 632 | 30 |
| | | | 700 | 0.8 | 0.7 | 27.5 | 402 | 10 |
| ponderosa pine (PP) | Cylindrical (Eq. 3) | 1.2 | 1025 | 6.1 | 1.5 | 83.5 | 727 | 60 |
| | | | 1010 | 3.5 | 1.2 | 66.7 | 712 | 50 |
| | | | 930 | 1.8 | 1.0 | 49.1 | 632 | 30 |
| | | | 700 | 0.7 | 0.7 | 29.6 | 402 | 10 |
| Douglas-fir (DF) | Cylindrical (Eq. 3) | 1.3 | 1025 | 6.5 | 1.5 | 80.0 | 727 | 60 |
| | | | 1010 | 3.8 | 1.2 | 63.8 | 712 | 45 |
| | | | 930 | 1.9 | 1.0 | 46.8 | 632 | 30 |
| | | | 700 | 0.8 | 0.7 | 28.1 | 402 | 10 |
| western larch (WL) | Cylindrical (Eq. 3) | 0.6 | 1025 | 3.0 | 1.1 | 129.9 | 727 | 95 |
| | | | 1010 | 1.7 | 0.9 | 105.7 | 712 | 75 |
| | | | 930 | 0.9 | 0.7 | 79.9 | 632 | 50 |
| | | | 700 | 0.3 | 0.6 | 50.1 | 402 | 25 |
| western red cedar (WC) | Cylindrical (Eq. 3) | 1.4 | 1025 | 6.8 | 1.6 | 78.3 | 727 | 55 |
| | | | 1010 | 3.9 | 1.3 | 62.3 | 712 | 45 |
| | | | 930 | 2.0 | 1.0 | 45.7 | 632 | 30 |
| | | | 700 | 0.8 | 0.7 | 27.4 | 402 | 10 |
| white spruce (WS) | Cylindrical (Eq. 3) | 2.0 | 1025 | 9.9 | 1.8 | 62.6 | 727 | 45 |
| | | | 1010 | 5.7 | 1.5 | 49.5 | 712 | 35 |
| | | | 930 | 2.9 | 1.1 | 35.8 | 632 | 20 |
| | | | 700 | 1.1 | 0.8 | 21.1 | 402 | 10 |
| pacific yew (PY) | Flat plate (Eq. 4) | 1.2 | 1025 | 6.1 | 0.8 | 42.1 | 727 | 30 |
| | | | 1010 | 3.5 | 0.5 | 29.7 | 712 | 20 |
| | | | 930 | 1.8 | 0.4 | 18.5 | 632 | 10 |
| | | | 700 | 0.7 | 0.2 | 8.5 | 402 | 5 |
| Sagebrush (SB) | Flat plate (Eq. 4) | 3.8 | 1025 | 18.8 | 1.5 | 27.1 | 727 | 20 |
| | | | 1010 | 10.9 | 1.1 | 19.1 | 712 | 15 |
| | | | 930 | 5.5 | 0.8 | 11.9 | 632 | 10 |
| | | | 700 | 2.2 | 0.4 | 5.5 | 402 | 5 |

## Acknowledgments

This research was funded by DOD/EPA/DOE Strategic Environmental Research and Development Program (SERDP) project number, RC19-1092.

**Nomenclature**

d        burner diameter (m).

*D*       needle diameter (mm)

ISI      Ignition sensitivity index (non-dimensional)

| | |
|---|---|
| $k_g$ | thermal conductivity of the gas (kW/m²K) |
| $k_s$ | thermal conductivity of the species (kW/m²K) |
| $h$ | convective heat coefficient (kW/m²K) |
| $T_f$ | flame temperature (K) |
| $T_m$ | mean flame temperature (K) |
| $T_s$ | initial surface temperature at ambient temperature (20ºC) |
| $t_{ig}$ | ignition time (s) |
| $q''_{conv}$ | convective heat flux (kW/m²) |
| $Pr$ | Prandtl number of the air |
| $Re$ | Reynolds number of gas flow around the needle/leaf |
| $L$ | projected width of the flat plate (i.e., leaf) perpendicular to the freestream (m) |
| $Nu$ | Nusselt number |
| $V_g$ | Bulk flow velocity (m/s) |
| $W$ | thickness of leaves (mm) |
| $w_{wet}$ | weight of live fuel as a wet sample (gr) |
| $w_{dry}$ | weight of a dried sample (gr) |
| $w_{cont}$ | weight of a container (gr) |

**Greek Symbol**

| | |
|---|---|
| $\Phi$ | equivalence ratio |

**Subscription**

| | |
|---|---|
| conv | Convective |
| cont | Container |
| f | flame |
| ig | ignition |
| g | Gas phase around the sample |
| H | Highest heat flux ($T_g$=1025K) |
| L | Lowest heat flux ($T_g$=700K) |
| S | Surface |
| s | species |

**Abbreviation**
25

| | |
|---|---|
| DF | Douglas-fir |
| FFB | flat flame burner |
| LFMC | live fuel moisture content |
| LP | longleaf pine |
| PP | ponderosa pine |
| PY | pacific yew |
| SB | Sagebrush |
| WC | western red cedar |
| WL | western larch |
| WS | white spruce |

**Abstract image**

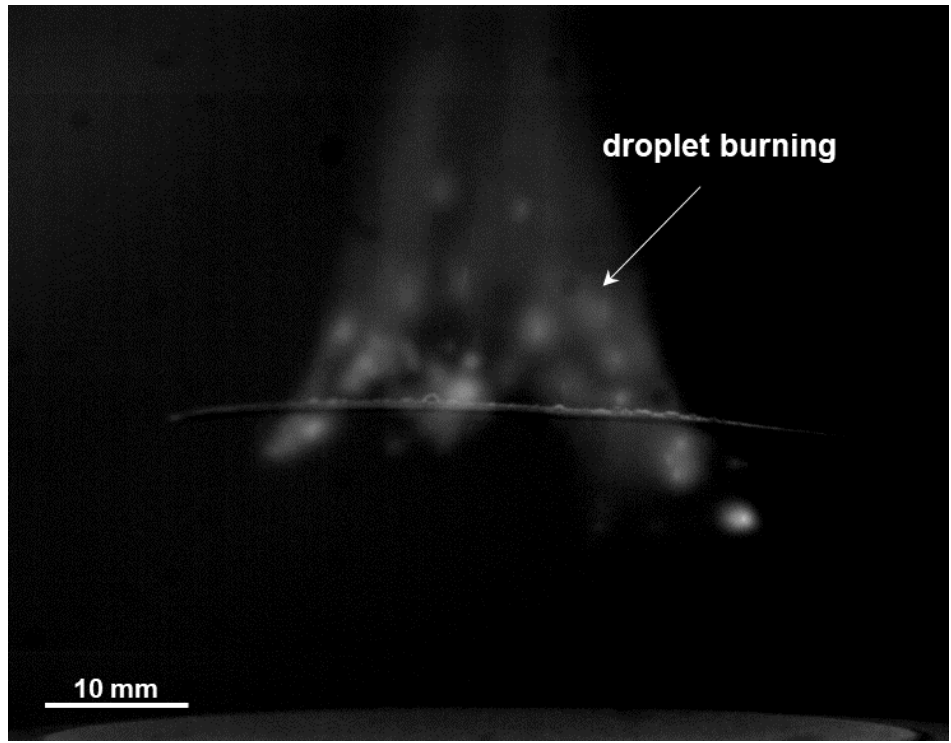

**Highlights:**

- This study seeks to understand the ignition and burning behavior of live, dried, and dead fuels for seven coniferous and one shrub species.
- Ejection and subsequent burning of droplets can occur prior to sustained flaming ignition only in live fuels and with sufficient heat flux.
- Ignition time of live fuels is much more sensitive to the heat flux than dead fuels.

Credit authorship contribution statement

Hamid Fazeli: Conceptualization, Methodology, Data curation, Software, Formal analysis, Investigation, Writing - original draft, Visualization.

William M. Jolly: Conceptualization, Methodology, Investigation, Writing review & editing, Co-Project administration, Funding acquisition.

David L. Blunck: Formal analysis, Methodology, Investigation, Writing - review & editing, Supervision, Project administration, Funding acquisition.29